\documentclass[11pt,a4paper]{article}
\usepackage[greek,english]{babel}
\usepackage{amssymb}
\usepackage{graphicx}
\usepackage{amsmath}
\usepackage{makeidx}
\usepackage{indentfirst}
\usepackage[T1]{fontenc}
\usepackage[utf8]{inputenc}
\usepackage{verbatim} 
\usepackage{authblk}

\newcounter{resultnum}[section]
\newcounter{conclusionnum}[section]
\newcounter{conditionnum}[section]
\newcounter{conjecturenum}[section]
\newcounter{examplenum}[section]
\newcounter{exercisenum}[section]
\newcounter{lemmanum}[section]
\newcounter{notationnum}[section]
\newcounter{theoremnum}[section]
\newcounter{definitionnum}[section]
\newcounter{corollarynum}[section]
\newcounter{remarknum}[section]
\newcounter{propositionnum}[section]
\newcounter{acknowledgementnum}[section]
\newcounter{algorithmnum}[section]
\newcounter{axiomnum}[section]
\newcounter{casenum}[section]
\newcounter{claimnum}[section]
\newcounter{summarynum}[section]
\newcounter{problemnum}[section]
\begin{document}

\title{Raychaudhuri equation in the Finsler-Randers spacetime and Generalized scalar-tensor theories}
\author{\vspace{.1 in} \textbf{Panayiotis C. Stavrinos} \\
{\small \textit{Department of Mathematics,National and Kapodistrian University of Athens}}\\
{\small \textit{Ilissia, Athens, 15784 Greece }}\\
{\small \textit{email: pstavrin@math.uoa.gr }}\\
\vspace{.1 in} \textbf{Maria Alexiou} \\
{\small \textit{Physics Division,National Technical University of Athens}}\\
{\small \textit{15780 Zografou Campus, Athens, Greece }}\\
{\small \textit{email: maria121289@hotmail.com  }}\\
}

\maketitle

\begin{abstract}
In this work, we obtain the Raychaudhuri equations for various types of Finsler spaces as the Finsler-Randers (FR) space-time and in a generalized geometrical structure of the space-time manifold which contains two fibres that represent two scalar fields $\phi^{(1)}, \phi^{(2)}.$We also derive the Klein-Gordon equation for this model. In addition, the energy-conditions are studied in a FR cosmology and are correlated with FRW model. Finally, we apply the Raychaudhuri equation for the model  $M \times \{ \phi^{(1)} \} \times \{ \phi^{(2)} \}$, where M is a FRW-spacetime.

\end{abstract}

\vskip3pt
\textbf{Keywords:}
Finsler Geometry,Raychaudhuri Equation,Scalar-Tensor Theories,

\vskip5pt

AMS: 53-XX, 83-XX, 85A40

\section{Introduction}

Raychaudhuri equation has been introduced by A.Raychaudhuri \cite{Ray}. It is a fundamental  equation in gravitation and cosmology and has been studied and generalized in many ways, in different cases \cite{ehlers,das,ellis,mohajan,hot,dad,kar1,kar2,lobo,k-ts,DDas,Abreu,mohseni1,Alba2,santos,ahmadi,moh,ra1,ra2,kar3}. In Finsler space-time this equation was introduced in \cite{st1}, \cite{st2}, \cite{idk1} and later studied in a different way \cite{Minguzzi} . Raychaudhuri equation describes the evolution of a gravitating fluid and provides a validation of our expectation that gravitation should be a universal attractive force between any two particles in general relativity.When cosmological fields such as dark energy,electromagnetic fields, spurionic fields or anisotropy are considered in the conventional Raychaudhuri equation,the evolution of the acceleration of the Universe is affected e.g. \cite{koivisto,st-k-st}. The form of the Raychaudhuri  equation varies when the metrical structure of the space changes, for example in spaces with torsion or generalized metric spaces such as a Finsler space time in which the Friedmann equations are modified \cite{st1}, \cite{st2},\cite{st-k-st}. In the case of null geodesics,the null Raychaudhuri equation, plays a key role in geometrical optics of a curved space-time. In addition, the Raychaudhuri equation ensures the existence of conjugate points with the famous singularity theorems in congruences and provides that the strong energy condition holds \cite{hawk}.The Raychaudhuri equation is produced by the structure of deviation equation of nearby geodesics or curves and it is considered as an index that records the evolution of the geodesics(focusing/defocusing) and plays a fundamental role in the dynamics of the fluid.The effects of gravity are encoded in the evolution of the expansion which is governed by Raychaudhuri's equations.  The different types of Raychaudhuri's equation are useful when we want to pass to different geometrical structures and phases during the evolution of the universe. All these types can describe the corresponding changes of the equations of motion. The Raychaudhuri equation  is related with the tidal force fields ($R_{jkl}^{i} \neq 0$), consequently, the gravitational field is transferred in it and the tidal field Ricci is included in the equation. When we consider a congruence of geodesics representing the motion of flow lines they pass through a surface which is vertical to the fluid .In this case, the accelerated expansion of the universe can be studied through its effects on a congruence of geodesics\cite{Alba1}. While the elementary particles are moving to adjacent geodesics of the fluid a deviation happens due to the gravitational field  and torsion in generalized metric spaces. The motion and the variation of the geodesic lines after some time causes a distortion of the fluid surface. Thus the variation of the volume (expansion/contraction) includes tidal field, rotation and shear which are connected to the deviation vector.Changes in the deviation vector between two nearby world flow lines are monitored by Raychaudhuri's equation, which is closely related to the evolution of the universe.

The deviation of geodesics is of fundamental significance in the general relativity and gravitation because it  interrelates  the interaction of curvature with the matter. The deviation of geodesics in generalized metric spaces of Finsler stucture has been studied by H.Rund  \cite{idk} and E.Cartan  \cite{cartan} and later in a series of papers for Finsler and Lagrange spaces and their applications in Finslerian space-time \cite{Asanov},\cite{st7},\cite{st5},\cite{st9}. 

Utilization of Finsler geometry in physical applications constitutes a natural metric generalization of the Riemannian space-time manifold. The Finsler geometry allows local anisotropies that intrinsically are included as vector variables $y^{i}=\frac{dx^{i}}{dt}$ in the framework of a tangent bundle or a higher order vector bundle of the manifold. By considering a different background we can  provide a useful information for a universe that includes a primordial vector field, for example an electromagnetic one or a scalar (inflaton) field, spinor field.

The Cartan's type torsion tensor characterizes all the geometrical concepts of Finsler geometry and contributes in its configuration as a physical geometry. Finsler-Cartan gravitational field theory is compatible with the structure of general relativity. 

A fundamental factor for the form of Raychaudhuri equation is obtained by considering the structure of the space and the kind of the curvature. Especially, in generalized metric spaces such as a Finsler spacetime where the curvatures are more than one.The concept of the Raychaudhuri equation is extended with the dependence of internal variables as the velocity(scalars, spinors or an anisotropic field) and extra terms.In that case, the Raychaudhuri equations are expressed with curvatures of the form $K_{jkl}^{i} (x,v),S_{\beta \gamma \delta}^{\alpha}(x,v)$ coupling with a direction $y$ or a velocity vector $v$. In a different way the authors in \cite{lobo} have studied a coupling between velocity and the Riemann curvature tensor. The dependence of curvatures on the direction (velocity) of a scalar fields is a consequence of the local anisotropy of the spacetime which is intrinsically considered in the space.So in an effective theory of modified gravity more general than the Einstein one, the geodesic deviation equation and the Raychaudhuri equation can include the effects of curvature-matter-velocity coupling. Furthermore, an anisotropic form of curvature can be considered for astrophysical considerations, for example, in a Finslerian cosmology \cite{basil2013}.

A type of Finsler space is the Finsler-Randers space hereafter (FR) \cite{rand}, which constitutes an important geometrical structure in Finsler spaces \cite{idk2},\cite{matsu} as far as its applications, in the general relativity and cosmology are concerned \cite{st-k-st},\cite{xinli}. In a FR space we can replace on the second part(one form) the electromagnetic potential with a quantity that represents a field of anisotropy, a scalar, or spinning particles. In the classical type of FR space-time the geometry contains information for the gravity and the electromagnetism. Of course that is not a complete and self-consistent theory of unification. However, we could not ignore the fact that it gives us a picture of what we should expect from a theory of unified field in a more extended geometrical framework.

The paper is organized as follows: In section 2 we give in brief some geometrical concepts from the theory of Finsler geometry. In section 3 we present the Raychaudhuri equation in the Finslerian spacetime and we give its form for a FR space. The energy conditions are examined in this model. In addition,  bounce conditions are presented and they are compared with the FRW cosmology. Finally in section 4 two forms of the Raychaudhuri equations are studied in a generalized scalar-tensor theory with scalars $\phi^{(1)}, \phi^{(2)}$  that play the role of fibres. Some concluding remarks are given in section 5.

\section{Preliminaries}

In the following, we briefly present fundamental geometrical concepts from the theory of Finsler spaces \cite{vsg}, \cite{idk3}.\\
We consider a smooth 4-dimensional manifold M, $ (TM, \pi , M)$, its tangent bundle with $\pi : TM \rightarrow M$ the projection. The set TM carries a natural differentiable structure induced by the base manifold M.\\

We also consider a local system of coordinates $x^{i}, i = 0, 1, 2, 3$ and U, a chart of M. Then the couple $(x^{i}, y^{a})$ is a local coordinate system on $\pi^{-1}(U)$ in TM, $\alpha=0,1,2,3$. A coordinate transformation on the total space TM is given by

\begin{equation}
\tilde{x}^{i} = \tilde{x}^{i} (x^{0},...,x^{3})
\end{equation}

\begin{equation}
det \Arrowvert\frac{\partial  \tilde{x}^{i}}{\partial x^{j}}\Arrowvert  \neq 0
\end{equation}

\begin{equation}
 \tilde{y}^{\alpha} = \frac{\partial  \tilde{x}^{\alpha}}{\partial x^{b}}y^{b}
\end{equation}

\begin{equation}
x^{\alpha}= \delta_{i}^{\alpha} x^{i}
\end{equation}

A Finsler metric on M is a function
F : TM $\rightarrow \mathbb {R}$ having the properties:\\

1. The restriction of F to $\widetilde{TM}=TM  \setminus \{0\}$ is of class $C^{\infty}$, and
F is only continuous on the image of the zero
cross-section in the tangent bundle to M.\\

2. F is positively homogeneous
of degree 1 with respect to $(y^{\alpha})$,
$F(x, ky) = kF(x, y), k \in \mathbb{R}_{+}^{*}.$\\

3. The quadratic form on $\mathbb{R}^{n}$  with the coefficients
$f_{ij} = \frac{1}{2} \frac{\partial^{2} F^{2}}{\partial y^{i} \partial y^{j}}$
defined on $\widetilde{TM}$ is non-degenerate $ (det(f_{ij}) \neq 0)$,
with rank $(f_{ij}) = 4$.

The existence of an horizontal distribution together with a vertical one determine a decomposition of the tangent bundle TTM into a Whitney sum

\begin{equation}
T_{(x, y)} (TM) = H_{(x, y)} \oplus V_{(x, y)} .
\end{equation}

A non-linear connection N on TM  can be defined on the total space of the tangent bundle and  defines a parallel transport on the base manifold.The coefficients of a non-linear connection are given by 
\begin{equation}
N_{j}^{a}= \frac {\partial G^{a}}{\partial y^{j}}
\end{equation}

where the coefficients $G^{a}$ are defined by
\begin{equation}
G^{a} = \frac{1}{4} f^{aj} (\frac {\partial^{2}F}{\partial y^{j}\partial x^{k}} y^{k} - \partial_{j} F),
\end{equation}

and the relation
\begin{equation}
\frac{dy^a}{ds} + 2G^{a}(x, y)=0
\end{equation}
follows from the Euler-Lagrange equations
\begin{equation}
\frac{d}{ds} (\frac {\partial F}{\partial y^{a}}) - \frac {\partial F}{\partial x^{a}} = 0.
\end{equation}

The transformation rule of the non-linear connection
coefficients is given by

\begin{equation}
\tilde {N_{i}^{a}}=\frac {\partial \tilde {x}^{a}}{\partial x^{b}} \frac { \partial x^{j}}{\partial \tilde {x}^{i}} N_{j}^{b}(x, y) + \frac {\partial \tilde {x}^{a}}{\partial x^{h}} \frac { \partial ^{2} x^{h}}{\partial \tilde{x}^{i} \partial \tilde{x}^{b}} y^{b}
\end{equation}

A local basis of $T_{(x,y)} (TM), (\delta_{i},\dot{\partial}_{\alpha})$ adapted to the
horizontal distribution N has the form

\begin{equation}
\delta_{i} = \frac{\delta}{\delta x^{i}}= \partial_{i} - N_{i}^{\alpha}(x,y) \dot{\partial}_{\alpha},
\end{equation}

where

\begin{equation}
\partial_{i} = \frac{\partial}{\partial x^{i}}, \hspace{3cm}
\dot{\partial}_{\alpha} = \frac{\partial}{\partial y^{\alpha}}
\end{equation}

and $N_{i}^{\alpha}(x, y)$ are the coefficients of the non-linear Cartan
connection N.
The concept of non-linear connection is fundamental
in the geometry of vector bundles and locally anisotropic
spaces. It is a powerful tool for a geometrical unification of fields.
For example, in the case of the gravitational field, the
non-linear connection in the framework of tangent bundle
unifies the external and internal spaces, i.e., the
position space (the base manifold M) with the tangent
space $T_{p} $M. In other words, it is connected with the local
anisotropic structure of space-time (depends on the
velocities).
The dual local basis is

\begin{equation}
\{ d^{i} = dx^{i}, \delta^{\alpha} = \delta y^{\alpha} = d y^{\alpha} + N_{j}^{\alpha} dx^{j} \}_{i,\alpha = \overline{0,3}}
\end{equation}

The transformation laws  for a local base and its dual are given by

\begin{equation}
\frac {\delta}{\delta \tilde{x}^{i}}=\frac{\partial x^{j}}{\partial \tilde {x}^{i}} \frac{\delta}{\delta x^{j}},  \hspace{3cm} \frac {\partial}{\partial \tilde{y}^{a}}=\frac{\partial x^{b}}{\partial \tilde {x}^{a}} \frac{\partial}{\partial y^{b}}
\end{equation}

\begin{equation}
d\tilde{x}^{i} = \frac{\partial \tilde{x}^{i}}{\partial x^{j}} dx^{j}, \hspace{3cm}
\delta \tilde{y}^{\alpha} = \frac{\partial \tilde{x}^{\alpha}}{\partial x^{b}} \delta y^{b}
\end{equation}

A d-connection on the tangent bundle TM of spacetime
is a linear connection on TM which preserves by
parallelism the horizontal distribution H and the vertical
distribution V on TM, rel. (5). A covariant derivative associated
with a d-connection becomes d-covariant. The tangent bundle is equipped with a (h-v) Sasaki-type metric including an horizontal and a vertical part and given by 

\begin{equation}
G= f_{ij} (x,y) dx^{i} \otimes dx^{j} + f_{ij} (x,y) \delta y^{i} \otimes \delta y^{j},
\end{equation}

where i,j=(1,..,4).

In gravitational metric theories on a space-time manifold as well on the tangent bundle metric-compatible connections play an important role to the parallel transport of vectors and to preserve their norm. In the framework of a tangent bundle, it is valid if the horizontal and vertical parts of the metric $f_{ij}$ are satisfied by the relations 

\begin{displaymath}
f_{ij|k}=0
\end{displaymath}
\begin{displaymath}
f_{ij}|_{\ell}=0
\end{displaymath} 

We consider a metrical d-connection $C \Gamma = (N_{j}^{\alpha},L_{jk}^{i},C_{jk}^{i})$ with the property

\begin{equation}
f_{ij|k}=\delta_{k}f_{ij} - L_{ik}^{h}f_{hj}-L_{jk}^{h}f_{ih}=0,
\end{equation}

\begin{equation}
f_{ij}|_{k}=\dot{\partial}_{k}f_{ij} - C_{ik}^{h}f_{hj}-C_{jk}^{h}f_{ih}=0
\end{equation}

where

\begin{equation}
L_{jk}^{i}=\frac{1}{2}f^{ir}(\delta_{j}f_{rk}+\delta_{k}f_{jr}-\delta_{r}f_{jk})
\end{equation}

\begin{equation}
C_{jk}^{i}=\frac{1}{2}f^{ir}(\dot{\partial}_{j}f_{rk}+\dot{\partial}_{k}f_{jr}-\dot{\partial}_{r}f_{jk}).
\end{equation}
In a Finsler gravitational theory, the Cartan connection preserves the rel. (17),(18) and it produces the Cartan-Finsler gravitational field theory convenient for a Finslerian relativity and cosmology.\\

The coordinate transformation of the objects $L_{jk}^{i}$ and $C_{jk}^{i}$ is

\begin{equation}
\tilde{L}_{jk}^{i}=\frac{\partial \tilde{x}^{i}}{\partial x^{h}} \frac{\partial x^{l}}{\partial \tilde{x}^{j}} \frac{\partial x^{r}}{\partial \tilde{x}^{k}} L_{ir}^{h}(x, y)+ \frac{\partial \tilde{x}^{i}}{\partial x^{r}} \frac{\partial^{2}x^{r}}{\partial \tilde{x}^{j} \partial \tilde{x}^{k}}
\end{equation}

\begin{equation}
\tilde{C}_{jk}^{i}=\frac{\partial \tilde{x}^{i}}{\partial x^{h}} \frac{\partial x^{l}}{\partial \tilde{x}^{j}} \frac{\partial x^{r}}{\partial \tilde{x}^{k}} C_{ir}^{h}(x, y)
\end{equation}

The Cartan torsion coefficients $C_{ijk}$ are given by
\begin{equation}
C_{ijk} = \frac{1}{2} \dot{\partial}_{k} f_{ij}
\end{equation}

while the Christoffel symbols of the first and second
kind for the metric $f_{ij}$ are

\begin{equation}
\gamma_{ijk}= \frac{1}{2} (\frac{\partial f_{kj}}{\partial x^{i}}+\frac{\partial f_{ik}}{\partial x^{j}}-\frac{\partial f_{ij}}{\partial x^{k}})
\end{equation}

\begin{equation}
\gamma_{ij}^{l}= \frac{1}{2} f^{lk} (\frac{\partial f_{kj}}{\partial x^{i}}+\frac{\partial f_{ik}}{\partial x^{j}}-\frac{\partial f_{ij}}{\partial x^{k}}),
\end{equation}

respectively. The torsions and curvatures which we use
are given by 

\begin{equation}
T_{kj}^{i} = 0, \hspace{3cm} S_{kj}^{i} =0
\end{equation}

\begin{equation}
R_{jk}^{i} = \delta_{k} N_{j}^{i} - \delta_{j} N_{k}^{i}, \hspace{3cm} P_{jk}^{i} = \dot{\partial}_{k} N_{j}{i} - L_{kj}^{i}
\end{equation}

\begin{equation}
P_{jk}^{i} = f^{im} P_{mjk}, \hspace{3cm} P_{ijk} = C_{ijk|l} y^{l}
\end{equation}

\begin{equation}
R_{jkl}^{i} = \delta_{l} L_{jk}^{i} + \delta_{k} L_{jl}^{i} +  L_{jk}^{h} L_{hl}^{i} -  L_{jl}^{h} L_{hk}^{i} +C_{jc}^{i} R_{kl}^{c}
\end{equation}

\begin{equation}
S_{jikh}=C_{iks} C_{jh}^{s} - C_{ihs} C_{jk}^{s}
\end{equation}

\begin{equation}
P_{ihkj} = C_{ijk|h} - C_{hjk|i} +C_{hj}^{r} C_{rik|l} y^{l} - C_{ij}^{r}C_{rkh|l} y^{l}
\end{equation}

\begin{equation}
S_{ikh}^{l} = f^{lj} S_{jikh}
\end{equation}

\begin{equation}
P_{ikh}^{l} = f^{lj} P_{jikh}
\end{equation}

The Ricci identities for the d-connection are

\begin{equation}
X^{i}|_{k}|_{h} - X^{i}|_{h}|_{k} =X^{r} R_{r}^{\hspace{0,1cm} i} \hspace{0,05cm}  _{kh} - X^{i}|_{r} R^{r} \hspace{0,05cm}  _{kh},
\end{equation}
\begin{equation}
X^{i}|_{k}|_{h} - X^{i}|_{h}|_{k} =X^{r} P_{r}^{\hspace{0,1cm} i} \hspace{0,05cm}  _{kh} - X^{i}|_{r} C^{r} \hspace{0,05cm}  _{kh} - X^{i}|_{r} P^{r} \hspace{0,05cm}_{kh}
\end{equation}
\begin{equation}
X^{i}|_{k}|_{h} - X^{i}|_{h}|_{k} = X^{r} S_{r}^{\hspace{0,1cm} i} \hspace{0,05cm}  _{kh}
\end{equation}

\section{Raychaudhuri Equation}

The Raychaudhuri equations are related with the kinematics of flows. Flows are generated by a vector field, they are the integral curves of a given vector field. These curves may be geodesics or non-geodesics.A  flow is a congruence of curves that are time-like, null or sometimes space-like. We are more interested  in deriving additional kinematic characteristics of such flows with extra terms. The evolution equations (along the flows) of quantities that characterize the flow in a given background spacetime are the Raychaudhuri equations. In fluid flows of cosmology there is a preferred 4-velocity vector field $u^{a} : u^{a} u_{a} = 1$ that represents the average motion of matter. Let $ \tau $ be the proper time along these world lines $ u^{a} = \frac{dx^{a}}{d \tau}$. The acceleration vector of $u^{\alpha}$ is  $\dot{u^{a}} = u^{a}_{; b} u^{b}$, which vanishes if and only if the flow lines are geodesics for a pseudo-Riemannian background.

The covariant differentation of the velocity field u is a second rank tensor
\begin{displaymath}
\nabla_{b} u_{a} = \sigma_{ab} + \omega_{ab} + \frac{1}{3} h_{ab} \Theta
\end{displaymath}

 which is splitted into three parts : the symmetric  traceless part, the shear of the flow 
\begin{displaymath}
\sigma_{ab} = \frac{1}{2} (\nabla_{b} u_{a}+ \nabla_{a} u_{b}) -(\frac{1}{3}) h_{ab} \Theta,
\end{displaymath}
the antisymmetric part, the rotation of the flow 
\begin{displaymath}
\omega_{ab}= \frac{1}{2} (\nabla_{b} u_{a} - \nabla_{a} u_{b}) 
\end{displaymath}
and the trace, the expansion of the flow 
\begin{displaymath}
\Theta = \nabla_{a} u^{a}.
\end{displaymath}

 The Raychaudhuri equation, giving the evolution of $\Theta$ along the fluid flow lines. Thus one has

 \begin{equation}
\dot{\Theta} + \frac{1}{3} \Theta^{2}  = 2 (\omega^{2} - \sigma^{2} ) + \dot{u}^{a}_{; a} - \frac{1}{2} \kappa (\mu + 3p) + \Lambda 
\end{equation}

This is the generic form of the Raychaudhuri equation, which is the fundamental equation of gravitational attraction and shows that shear, energy density and pressure tend to make matter collapse or decelerate the expansion while vorticity and a positive cosmological constant tend to make matter expand.

\subsection{Raychaudhuri Equation in a Finsler Space-time}

 \textbf{Finslerian congruences}\\
 \noindent 
Suppose $(F^{4},f_{ij}(x,y))$ is a four dimensional differentiable
manifold and $f_{ij}(x,y)$ the anisotropic Finslerian metric is
assumed to have signature $(+,-,-,-)$ for any $(x,y)$.

The motion of a particle in a Finslerian space-time $F^{4}$ is
described by a pair $(x^{i},u^{i})$ where $x^{i} \in F^{4}$ and
$u^{i}=\dfrac{dx^{i}}{d\tau},i=1.2.3.4$  the 4-velocity of the particle (time-like/null) ($\tau$ is
proper time) which  represents the tangent of its world-line
expressing the motion of typical observers in the Finslerian locally
anisotropic universe.

A \emph{smooth congruence} in an open coordinate neighborhood $U$
of $F^{4}$ can be represented by a preferred family of world lines
(time-like curves or null) such that through each couple $(x,u)\in U$
there passes precisely one curve in this family in which $u$ is
the tangent vector of this curve to that point $x$. This
consideration is analogous to the Riemannian context.In the framework of a tangent bundle the extended congruence has the form $(x^{(i)}(t),y^{(i)}(t))$.The dependence on a line-element ($x,y$) can also be considered in this approach \cite{stkst1}. 

The metric of pseudo-Finslerian space-time is described by the relation
\[ ds^2=F^2(x,y)=f_{ij}y^ iy^ j\]

The time-like, null and space-like curves can be defined in the
Finslerian framework by the following relations \cite{Ishi}
\begin{align}
    & \text{time-like} & & f_{ij}(x,y)u^i u^j >0\notag\\
    & \text{null-like} & & f_{ij}(x,y)u^i u^j =0\\
    & \text{space-like} & & f_{ij}(x,y)u^i u^j<0\notag
\end{align}

In the framework of of Finsler cosmology we have also considered the form of Riemannian osculation of a metric  \cite{st-k-st}. The Finslerian metric tensor and a contravariant vector field $y^{i}(x)$ may be used to construct the Riemannian metric tensor $a_{ij}(x)=g_{ij}(x,y(x))$. The Riemannian space associated with this metric tensor is called the osculating Riemannian space.This gives us the possibility to view some cosmological considerations in a 4-dim space-time framework.
 In the following we present the content of the Raychaudhuri Equations for a Finsler space-time.We assume Finslerian fluid congruences that the matter flow
lines of the  fluid are time-like geodesics and are parameterized
by the proper time $\tau$ so that a vector field $u^i (x)$ of
tangents is normalized to the unit length $u^i =\dfrac{y^i}{F}$.

Using the $\delta$-differentiation in the direction of $u^i (x)$ for
a  congruence of fluid lines (not necessarily geodesics) 
the expansion,  vorticity and the shear are defined by the
following forms: \cite{st2}
\begin{align}   
    & \widetilde{\varTheta}
    =Z_{ij}h^{ij}=u^{i}_{|i}-C^{i}_{im}\dot{u}^m\\
    &\tilde{\omega}_{ik}=Z_{[ik]}+\dot{u}_i u_k -\dot{u}_k u_i \\
    &\tilde{\sigma}_{ik}=Z_{(ik)}-\frac{1}{3}\widetilde{\varTheta}
    h_{ik}-2C_{ikm}u^m -\dot{u}_iu_k-\dot{u}_ku_i
\end{align}
where $\dot{u}^i =u^{i}_{|k}u^k =Z^{i}_{k}u^k$ and ``$|$''
denotes the Riemannian covariant derivative associated with the
osculating Riemannian metric $a_{ij}(x)=g_{ij}(x,u(x))$. The
symbols ``$[\;]$'', ``$(\;)$'' denote the antisymmetrization and
symmetrization of $Z_{ik}$ respectively. The tensor
$C_{ijk}= \frac{1}{2} \frac{\partial f_{ij}(x,y)}{\partial y^k}$
is symmetric in all its subscripts.

Therefore the  extended Finslerian
covariant derivative of $u$ can be expressed by
\begin{align}   
    Z_{ik}=\frac{1}{3}\widetilde{\varTheta}h_{ik}+\tilde{\sigma}_{ik}+\tilde{\omega}_{ik}+\dot{u}_i
    u_k
\end{align}

The consideration of a Finslerian incoherent fluid implies that
the fluid lines are geodesics and $\dot{u}^i =Z^{i}_{k}u^k =0$.
In this case the Finslerian geodesics coincide with the Riemannian
ones of a $u$-Riemannian space (osculating Riemannian).

The commutation relations of
$\delta$-covariant derivative of the vector field $u^i (x)$ gives
\begin{align}   
    u_{i;hk}-u_{i;kh}=L^{j}_{ikh}u_j
\end{align}
where $L^{i}_{jhk}$ curvature tensor is derived by the
$\delta$-\emph{covariant derivative with respect to the} osculating
affine connection coefficients $a^{i}_{jk}(x,u(x))$ [30].
\begin{displaymath} \notag
L^i_{jhk}\left(x,u(x)\right)= \left( \frac{\partial
L^i_{jh}}{\partial x^k}+\frac{\partial L^i_{jh}}{\partial u^l} \frac{\partial
u^l}{\partial x^k} \right)- \left( \frac{\partial L^i_{jk}}{\partial x^h}+ \frac{\partial
L^i_{jk}}{\partial u^l} \frac{\partial u^l}{\partial x^h} \right)+
L^i_{mk}L^m_{jh}-L^i_{mh}L^m_{jk}
\end{displaymath}

 We finally obtain
\begin{align}   
    \frac{d\widetilde{\varTheta}}{d\tau}=-\frac{1}{3}\widetilde{\varTheta}^2
    -\tilde{\sigma}_{ik} \tilde{\sigma}^{ik}+\tilde{\omega}_{ik}\tilde{\omega}^{ik}-L_{i\ell}u^i
    u^\ell +\dot{u}^{i}_{;i}
\end{align}
This is {\em Raychaudhuri's } equation of the Finslerian
space-time.%
The change of expansion which is expressed by
$\dfrac{d\widetilde{\varTheta}}{d\tau}$ depends on the
$V$-anisotropic
behavior of tensor $C^{i}_{jk}$ along the matter flow lines.%
When we consider an incoherent fluid, the fluid-lines are
geodesics and the last term of right hand side of (44) is
$\dot{u}^i =0$. In this case the Raychauduri equation is reduced
to the form of a $u$-Riemannian metric of osculating space associated with the
congruence of geodesics.

A perfect fluid in the Finslerian space time case has the osculating form
\begin{align}   
    T_{ij}(x,u(x))=(\mu +p)u_i (x)u_j (x)+pa_{ij}
\end{align}
where $p=p(x)$, $\mu =\mu (x)$ and $u_{i}(x)$ represent the pressure, the
density of the fluid and the fluid-four velocity respectively.

The Einstein equations can be written in the form
\begin{align}   
    L_{ij}(x,u(x))=K\left( T_{ij}(x,u(x))-
    \frac{1}{2}T^{k}_{k}a_{ij}\right) , ~~~ K:\text{constant}
\end{align}
where the Ricci tensor $L_{ij}$ is directly determined by the
matter energy-momentum tensor $T_{ij}$ and at each point associated
with the osculating Riemannian metric tensor
$a_{ij}(x)=g_{ij}(x,u(x))$. Substitution of (45) to (46) gives
\begin{align}\label{18}   
    L_{ij}u^i u^j =\frac{1}{2} K(\mu +3p)
\end{align}
The term $L_{i\ell}u^i u^\ell$ corresponds to an anisotropic
gravitational influence of the matter along the world lines of the
fluid and it expresses the tidal force of the field.

From rel.(45) the Raychauduri equation in the case of a perfect fluid
 is given  by
\begin{equation}
\dot{\tilde{\Theta}}=\frac{d\tilde{\Theta}}{d \tau}=-\frac{1}{3}
\tilde{\Theta}^2-\tilde{\sigma}_{ik}\tilde{\sigma}^{ik}+
\tilde{\omega}_{ik}\tilde{\omega}^{ik} -\frac{1}{2} K(\mu+3 p) +
\dot{u}^i_{;i}
\end{equation}

The relation
\begin{align}\label{19}   
    L_{i\ell}u^i u^\ell >0
\end{align}
implies the so called strong energy condition for every
time-like vector $u^{\alpha}$ tangent to time-like geodesics.
 We notice that the fluid energy $\mu$ and pressure $p$
satisfy the energy condition $\mu +p >0$. This condition uniquely
defines the Finslerian world lines (congruences) of the fluid with
$u(x)$ tangent vector field analogous to that of Riemannian
framework \cite{bruni} in a region of Finsler space-time which is called "osculating Riemannian manifold" \cite{idk}.
 The term $L_{il}u^i u^l>0$ (rel.46) can also  be considered as a key
for the existence of conjugate points in the Finslerian space-time
structure.\\

\subsubsection{Raychaudhuri Equation in a Finsler-Randers (FR) space}
\textbf{The geodesics}

The Lagrangian metric function in a FR space is given by 
\begin{equation} \label{r14}
\mathfrak{L}=(g_{ij}\dot{x}^{i}\dot{x}^{j})^{1/2}+kA_{i}\dot{x}^{i}, k= constant
\end{equation}
where $A_{i}(x)$ is the electromagnetic potential and  $g_{ij}(x)$ the Riemannian gravitational potential with $\dot{x}^{i} = \frac{dx^{i}}{ds}$.

From (19) we get 

\begin{equation} \label{r15}
\frac{\partial \mathfrak{L}}{\partial x^{a}}= \frac{1}{2 (g_{ij}\dot{x}^{i}\dot{x}^{j})^{1/2} } \frac{ 
\partial g_{ij} }{ \partial x^{a} } \dot{x}^{i}\dot{x}^{j}+k \frac{\partial A_{i}}{\partial x^{a}} \dot{x}^{i}
\end{equation}

\begin{displaymath} \label{r16}
\frac{\partial \mathfrak{L}}{\partial \dot{x}^{a}}= \frac{1}{ 
(g_{ij}\dot{x}^{i}\dot{x}^{j})^{1/2}} g_{aj}\dot{x}^{j}+kA_{a}
\end{displaymath}

Thus the  Euler--Lagrange equations are derived by the relation 
\begin{equation} \label{r18}
\frac{d}{ds} \left(\frac{\partial \mathfrak{L}}{\partial \dot{x}^{a}}\right)-
\frac{\partial \mathfrak{L}}{\partial x^{a}}=0
\end{equation}

The last equation gives us
\begin{equation} \label{r20}
\frac{d\dot{x}^{i}}{ds}+\Gamma^{i}_{mn}\dot{x}^{m}\dot{x}^{n}+k 
(g_{ij}\dot{x}^{i}\dot{x}^{j})^{1/2}F^{i}_{j}\dot{x}^{j}=0
\end{equation}

$F_{ik}$ represents the electromagnetic field. Eq.\eqref{r20} coincides with the equation of motion of a charged particle in a gravitational field 
\begin{equation}
\frac{d\dot{x}^{i}}{ds}+\Gamma^{i}_{mn}\dot{x}^{m}\dot{x}^{n}=-k F^{i}_{j}\dot{x}^{j} 
\end{equation}
or for a vector V parallel to $\dot{x}^{m}$ we get
\begin{equation}
\frac{dV^{i}}{d \tau} + \Gamma_{mn}^{i}V^{m}V^{n} = -k F_{j}^{i} V^{j},
\end{equation}
with
\begin{equation}
 \dot{V}^{i} = V_{;k}^{i} V^{k}
\end{equation}

In a FR space the quantity $\dot{V}_{; i}^{i} \neq 0$ applies in geodesics because the equation of geodesics  is the Lorentz  force (rel.53). It is encorporated into the Raychaudhuri Equation generalizing the case of the Riemannian ansantz in which the Lorentz Force is introduced in the Raychaudhuri Equation only  for non-geodesics flows in the presence of an electromagnetic field \cite{hot},\cite{k-ts}. Moreover, in a general Finsler spacetime of an osculating Riemannian form the flow lines follow geodesics motion when the acceleration is equal to zero.\\

In general, we can  consider a vector $V^{\ell}(x)$ tangent to the flow lines in FR space, then the equation of geodesics is given by
\begin{equation}
\frac{d^{2}V^{l}}{ds^{2}} + \Gamma^{l}_{ij}y_{i}y_{j}+ g^{lm}(\partial_{j}V_{m} - \partial_{m}V_{j})y^{j}=0 
\end{equation}
where $ \Gamma_{ij}^{l}$ represent the Christoffel symbols and $g^{lm}$ the components of the inverse metric of the Riemmanian space-time, $V_{m} = \phi(x) \widehat{V_{m}}$ with $ \widehat{V_{m}}$ the unit vector in the direction of $V_{m}$ and $\phi(x)$ a scalar. We observe that in the equation of geodesics we have an additional term, $g^{lm}(\partial_{j} V_{m}) - \partial_{m}V_{j}))y^{j}$, which contains a rotation.For the case of electromagnetic waves we must modify the above equation by using an affine parameter $\lambda$.This is because the world line of an EM wave is null. In geometric optics the direction of propagation of a light ray is determined by the wave vector tangent to the ray.Thus we have:
\begin{equation}
\frac{d \stackrel{w}{V^{l}}}{d \lambda} + \Gamma^{l}_{ij}\stackrel{w}{V_{i}}\stackrel{w}{V_{j}}+g^{lm}(\partial_{j}\stackrel{w}{V_{m}} - \partial_{m}\stackrel{w}{{V}_{j}})\stackrel{w}{V^{j}}=0
\end{equation}

If we substitute the relation $V_{i;j}-V_{j;i}=V_{i,j}-V_{j,i}$ from (57) with the vorticity  $\widetilde{\omega}_{ij}$ we obtain
\begin{equation}
\frac{d y^{\ell}}{ds} +\Gamma^{l}_{ij}y_{i}y_{j}+g^{li} \widetilde{\omega}_{ij}y^{j}=0, y^{i}=\frac{dx^{i}}{dt}
\end{equation}

We note from (57) that the y-dependence of the metric is a consequence of the existence of the anisotropic field $V_{k}$ in a FR space and the term of vorticity can be included in the equation of geodesics.(rel.59).

\textbf{The Raychaudhuri equation}\\
Taking into account the presence of an electromagnetic field $F_{ik}$ from rel.(55) the kinematical parameters $\widetilde{\Theta}, \widetilde{\omega},\widetilde{\sigma}$ of (39)-(41) are modified  in the form
\begin{align}   
    & \widetilde{\varTheta}=V^{i}_{|i}+f^{m} C^{i}_{im}\\
    &\widetilde{\sigma}_{ik}=Z_{(ik)}-\frac{1}{3}\widetilde{\varTheta}
    h_{ik}-2C_{ikm}V^m+\Gamma_{ki0}-\Gamma_{i0k}+2kF_{ki}\\
    &\widetilde{\omega}_{ik}=Z_{[ik]}+\Gamma_{ki0}-\Gamma_{i0k}+2kF_{ki}.
\end{align}

we put $f^{m}=\Gamma_{jk}^{m}V^{j}V^{k}+kF_{j}^{m}V^{j}$ and $\Gamma_{i0k}=\Gamma_{ijk} V^{j}$
Here "|" denotes the osculating covariant derivative.
In the above equations we used the rel. (55) which is the Lorentz force law in the FR space 
\hspace{5cm}

By using Cartan's covariant differentiation we get the Cartan curvature tensor $\tilde{R}_{jkl}^{i}$ and the Ricci tensor $\tilde{R}_{jk}$. In the framework of a tangent bundle the Raychaudhuri equation has been given in \cite{st1}. We can derive the Raychaudhuri equation  for  the model of FR spacetime (50) in the Finslerian  tangent bundle where instead  of an electromagnetic potential we can consider a time-like vector field $X^{n}$.
\begin{displaymath}
X^{m}\widetilde{\Theta}_{|_{m}}=\frac{d \widetilde{\Theta}}{d\tau}=\tilde{R}_{lm}X^{l}X^{m} - T_{mk}^{i}X_{|i}^{k} X^{m} - \widetilde{R}_{mk}^{b}X^{k}_{|_{b}} X^{m} - X^{m}_{|_{l}}X^{l}_{|_{m}}=
\end{displaymath}
\begin{equation}
=\widetilde{R}_{km}X^{k}X^{m}-T_{mk}^{i}(\frac{1}{3} \widetilde{\Theta} \tilde{h}_{i}^{k} + \widetilde{\sigma}_{i}^{k} + \widetilde{\omega}_{i}^{k}) X^{m} - \widetilde{R}_{mk}^{b} (\frac{1}{3} \widetilde{\Theta} \tilde{h}_{b}^{k} + \sigma_{b}^{k} + \omega_{b}^{k}) X^{m} - \frac{1}{3} \widetilde{\Theta}^{2} - \sigma_{l}^{m} \sigma_{m}^{l} - \omega_{l}^{m} \omega_{m}^{l}
\end{equation}

By considering the rel.(26) and the integrability condition  $\widetilde{R}_{mk}^{b}=0$ of rel.(27) we get from (63) the equation

\begin{equation}
X^{m}\widetilde{\Theta}_{|_{m}}=\widetilde{R}_{km}X^{k}X^{m}- \sigma_{l}^{m} \sigma_{m}^{l} - \omega_{l}^{m} \omega_{m}^{l}-\frac{1}{3} \widetilde{\Theta}^{2}
\end{equation}
Taking into account the form of Ricci tensor $\widetilde{R}_{kj}$ in FR space  \cite{matsu},\cite{yasuda} we get
\begin{equation}
\widetilde{R}_{kj}=R_{kj}+\Delta_{00k|j}-\Delta_{00j|k}+ \Delta_{km0} \Delta_{0j}^{m}- \Delta_{kmj} \Delta_{00}^{m}
\end{equation}
where $\Delta_{mj}^{i}$ represents the deformation tensor (see Appendix A)
\begin{equation}
\Delta_{hmj}=f_{ih} \Delta_{mj}^{i}
\end{equation}
$R_{kj}$ is the Ricci tensor of the Riemannian space and $f_{ih}$  represents the metric of FR spacetime. After the rel.(65) the Raychaudhuri equation for the horizontal subspace of the tangent bundle is written
\begin{equation}
X^{j}\tilde{\Theta} |_{j}=-\frac{1}{3} \tilde{\Theta}^{2} +\{R_{kj}+\Delta_{00k|j}-\Delta_{00j|k}+ \Delta_{km0} \Delta_{0j}^{m}- \Delta_{kmj} \Delta_{00}^{m}\} X^{k}X^{j}- \sigma_{l}^{j} \sigma_{j}^{l} - \omega_{l}^{j} \omega_{j}^{l}
\end{equation}

The form of Raychaudhuri's equation in the vertical space of a FR space can be given by\\
\begin{displaymath}
X^{g}\Theta |_{g}=\frac{d\Theta}{d\tau}=-\frac{1}{3} \tilde{\Theta}^{2}+S_{hg}X^{h}X^{g} - S_{ge}^{h}X^{e} |_{h}X^{g} - X^{g} |_{f}X^{f}|_{g}=
\end{displaymath}
\begin{equation}
 = -\frac{1}{3}\Theta^{2}-\sigma_{f}^{e}\sigma_{e}^{f}- \omega_{f}^{e}\omega_{e}^{f} + S_{hg}X^{h}X^{g} - S_{ge}^{f}X^{g} (\frac{1}{3}\Theta h_{f}^{e} + \sigma_{f}^{e} + \omega_{f}^{e}).
\end{equation}
The symbols "$_{|}$", "$|$" denote the horizontal and vertical covariant derivatives ref.\cite{idk3}, \cite{vsg}.
By imposing $S_{ge}^{f} = 0$ from  the rel(26),  the S-Ricci curvature of the FR  space gives \cite{st3} \\
\begin{displaymath}
S_{hg}=\frac{\phi^{2}}{2F^{2} \sigma^{2}} \left[ \frac{3 \sigma^{2} (y_{h} y_{g} - g_{hg}) -\beta^{2}(4 y_{h} y_{g} +3 g_{hg})}{2} +\beta \mathcal{S}_{hg}(y_{h}\widehat{V_{g}}) \right]      
\end{displaymath}
where $\mathcal{S}_{hg}$ is an operator and denotes symmetrization of the indices h,g we also considered that $\widehat{V}_{a} \widehat{V}^{a} =1$ for timelike vectors and $\sigma=(g_{ij}y^{i}y^{j})^{1/2}$,$\beta= y^{a}\widehat{V}_{a}$.
Therefore the Raychaudhuri equation in a FR-space takes the form\\

\begin{equation}
\frac{d \Theta}{d \tau}=-\frac{1}{3}\Theta^{2}-\sigma_{f}^{e}\sigma_{e}^{f}- \omega_{f}^{e}\omega_{e}^{f} + \{ \frac{\phi^{2}}{2F^{2} \sigma^{2}} \left[ \frac{3 \sigma^{2} (y_{h} y_{g} - g_{hg}) -\beta^{2}(4 y_{h} y_{g} +3 g_{hg})}{2} +\beta \mathcal{S}_{hg}(y_{h}\widehat{V_{g}}) \right]                                                               \}X^{h}X^{g}
\end{equation}
The form of Raychaudhuri equation in the vertical space can be interpreted as an internal anisotropic contribution in the evolution of the universe.

\subsection{Energy Conditions in a FR Space}

Energy conditions are of great significance in cosmology because in relation with the Friedmann and the Raychaudhuri equations constitute a key role in universe's evolution \cite{hawk}.In the following we give the energy conditions in a FR space. In this space an important role is played by the scalar of anisotropy $Z_{t}$.The Friedmann-like equations of the generalized form of the FR-type cosmology have been studied in \cite{st-k-st}. The form of these equations are given by the relations

\begin{equation}
\frac{\ddot{a}}{a}  + \frac{3}{4} \frac{\dot{a}}{a} \dot{u_{0}} = -\frac{4 \pi G}{3} \left( \mu + 3P \right) 
\end{equation}
\begin{equation}
 \frac{\ddot{a}}{a} + 2 \frac{\dot{a}^{2}}{a^{2}} + 2 \frac{k}{a^{2}} + \frac{11}{4}  \frac{\dot{a}}{a} \dot{u_{0}} = 4 \pi G ( \mu - P) 
\end{equation}
From rel.(70) and (71) we obtain the Friedmann-like equation
\begin{equation}
\left(\frac{\dot{a}}{a} \right)^{2}+\frac{\dot{a}}{a}Z_{t}=\frac{8 \pi G}{3} \mu- \frac{k}{a^{2}}
\end{equation}
where we set $u_{0}(t) = \phi(x) \hat{u_{0}}$,with  the time component of the unit vector $\hat{u_{a}}$ and $Z_{t}$  defined as $Z_{t}=\dot{u_{0}}$.
Taking into account relations (70)-(72) we obtain
\begin{equation}
\mu = \frac{3}{8 \pi G} \left( \frac{\dot{a^{2}}}{a^{2}} + \frac{\dot{a}}{a} Z_{t} + \frac{k}{a^{2}}  \right)=\frac{3}{8\pi G} (H^{2} + \frac{k}{a^{2}}+ Z_{t}H)
\end{equation}
with $H=\frac{\dot{a}}{a}$. From (73) in order to be valid  $\mu \geq 0$ the scalar should be $Z_{t} \geq 0$. We also have
\begin{equation}
P=-\frac{1}{8\pi G} \left[2 \left(\frac{\ddot{a}}{a}\right)+H^{2} + \left(\frac{k}{a^{2}} \right) +\frac{5Z_{t}}{2} H \right]
\end{equation}
From(70)-(74) we obtain
\begin{equation}
\mu+P=\frac{1}{8\pi G}[-2\frac{\ddot{a}}{a}+2H^{2}+2\frac{k}{a^{2}}+\frac{1}{2}Z_{t}H]
\end{equation}
and
\begin{equation}
\mu + 3P = \frac{-3}{4 \pi G} \left[ \frac{\ddot{a}}{a} + \frac{3}{4} \frac{\dot{a}}{a}Z_{t}\right]
\end{equation}
The strong energy condition $\mu + 3P \geq 0$ is obtained for $\frac{\ddot{a}}{\dot{a}} \leq -\frac{3}{4} Z_{t}$ consequently $\ddot{a} < 0 (Z_{t} >0) $.We also have $\mu+P  \geq 0$  if $\ddot{a} < 0$ from (75). In addition the weak energy conditions (WEC) and the null energy conditions (SEC) are also satisfied.

A bounce occurs in the universe when WEC,NEC,SEC are violated for a short interval of a point time (bounce time) with $\dot{a}=0$ and $\ddot{a} > 0$ \cite{singhsingh}, \cite{singsing}, \cite{singggsinggg}. In analogy with FRW-universe for a FR-cosmology a bounce can be considered if the above mentioned conditions for the WEC and SEC are violated. In this case $HZ_{t}=0$ and $\ddot{a}>0$. In a bounce time the SEC conditions of a FR spacetime and of a FRW-cosmology one are identified.We notice that the bounce conditions $\mu +P < 0, \mu +3P <0$ are equivalent to
\begin{equation}
2(H^{2}+ \frac{k}{a^{2}})+ \frac{1}{2} H Z_{t} < 2 \frac{\ddot{a}}{a} \Rightarrow \ddot{a}>0
\end{equation}
\begin{equation}
\ddot{a} > - \frac{3}{4}Z_{t} \dot{a} \Rightarrow \ddot{a} >0
\end{equation}

\section{ Raychaudhuri Equation in Generalized Scalar-Tensor Theory} 
\subsection{The model}

Scalar-tensor theories constitute a fundamental part of studying general relativity and cosmology. The significance of scalar-tensor and Brans-Dicke theories to the cosmology has mainly pointed out in \cite{far},\cite{bamba} et all. On the other hand, the simplest standard models of inflation involve one scalar field. However in string or supergravity theories it is possible to study models with several different scalar fields especially if they have some new properties. The mechanism that two scalars $\sigma^{(1)},\sigma^{(2)}$ drive the inflation in a more complicated theory is useful for studying the evolution of the universe. When the standard mechanism does not work, two scalars can give us an additional freedom in finding realistic models of inflationary cosmology\cite{Linde}.\\

In the framework of generalized metric structures of  scalar-tensor theory of gravitation a Langrangian density has been studied \cite{stik}. Furthermore, the field equations for the Finslerian gravitational field in a space of the form $M\times \{\phi^{(1)}\}\times \{\phi^{(2)}\}$ are derived, where M represents a pseudo-Riemannian space-time manifold with $\phi^{(1)},\phi^{(2)}$ two non-commutative G-numbers(Grassmannian).In our case we consider a 4-pseudo-Riemannian manifold coupled with two scalars $\phi^{(1)}, \phi^{(2)}$ playing the role of fibres or the internal variables. Physically these can represent e.g. the inflaton and an anisotropy come from a scalar field.\cite{koivisto}

In this space the adapted frame has the form
$\frac{\partial}{\partial Z^{M}} \equiv \left( \frac{\delta}{\delta x^{\alpha}} = \frac{\partial}{\partial x^{\alpha}} - N_{\alpha}^{(1)} \frac{\partial}{\partial \phi^{(1)}} - N_{\alpha}^{(2)} \frac{\partial}{\partial \phi^{(2)}},\frac{\partial}{\partial \phi^{(1)}},\frac{\partial}{\partial \phi^{(2)}} \right)$\\

\begin{displaymath}
dZ^{M} \equiv (dx^{\alpha}, \delta \phi^{(1)} = d \phi^{(1)} + N_{\alpha}^{(1)} dx^{\alpha}, \delta \phi^{(2)} = d\phi^{(2)} + N_{\alpha}^{(2)} dx^{\alpha})
\end{displaymath}
We put 
\begin{displaymath}
X_{\alpha} = \frac{\delta}{\delta x^{\alpha}},  X_{(1)}=\frac{\partial}{\partial \phi^{(1)}}, X_{(2)} = \frac{\partial}{\partial \phi^{(2)}}
\end{displaymath}
and
\begin{displaymath}
X_{M} =( {X_{\alpha},X_{(1)},X_{(2)}}),
\end{displaymath}
with 
\begin{equation}
\frac{\partial}{\partial \phi^{(1)}} \cdot \frac{\partial}{\partial \phi^{(1)}} =g_{(1)(1)}=g_{(2)(2)}=\frac{\partial}{\partial \phi^{(2)}} \cdot \frac{\partial}{\partial \phi^{(2)}}=\phi(x^{\alpha})
\end{equation}
where $\phi$ is a  scalar $\alpha=0,1,2,3$.

The geometrical concepts $N_{a}^{(1)},N_{a}^{(2)}$ represent the non-linear connections of scalar fields $\phi^{(1)},\phi^{(2)}$ with respect to the space-time coordinates.\\
The inverse metric of $G_{MN}$ is given by $G^{MN} = [g^{\alpha \beta},g^{(1)(1)},g^{(2)(2)}]$ where $g^{\alpha \beta}$ is the inverse metric of $g_{\alpha \beta}$ for M. The covector $X_{M}$ has the inverse $X^{N}= G^{MN}X_{M}$.
\begin{displaymath}
X^{(1)}=\phi^{-1} X_{(1)}=g^{(1) (1)} X_{1}
\end{displaymath}
\begin{displaymath}
X^{(2)}=\phi^{-1} X_{(2)}=g^{(2)(2)}X_{(2)}
\end{displaymath}
\begin{displaymath}
\phi^{-1}=g^{(1)(1)}=g^{(2)(2)}
\end{displaymath}

The metric structure of this model concerning the adapted basis is given by 
\begin{equation}
G= G_{MN}dZ^{M}dZ^{N} = g_{\alpha \beta}dx^{\alpha}\otimes dx^{\beta} + g_{(1)(1)} \delta \phi^{(1)} \otimes \delta \phi^{(1)} +  g_{(2)(2)} \delta \phi^{(2)} \otimes \delta \phi^{(2)}
\end{equation}
\\
where $G_{MN}=\{g_{\beta \gamma}(x^{\alpha}), g_{(1)(1)}(x^{\alpha}), g_{(2)(2)}(x^{\alpha}) \} $ with $ g_{(1)(1)}(x^{\alpha})=g_{(2)(2)} (x^{\alpha})=\phi(x^{a}) \neq 0 $ is a scalar function.The curvature tensor in this space is written

\begin{equation}
R_{LNM}^K=X_{M} \Gamma_{LN}^{K}-X_{N} \Gamma_{LM}^{K}+ \Gamma_{LN}^{Z} \Gamma_{ZM}^{K}- \Gamma_{LM}^{Z} \Gamma_{ZN}^{K} +  \Gamma_{LZ}^{K} W_{NM}^{Z} 
\end{equation}

In our approach,we shall adopt the case where all the coefficients of torsion tensors are equal to zero and $ g_{a(1)}=g_{(1)a}=g_{a(2)}=g_{(2)a}=g_{(1)(2)}=g_{(2)(1)}=0$. The Ricci tensor in this model is given by

\begin{equation}
R_{MN}\equiv R_{MLN}^{L} \equiv( R_{\alpha \beta},..., R_{(1)(1)}, R_{(2)(2)})
\end{equation}
or
\begin{equation}
R_{MN}=\{ R_{MN \alpha}^{\alpha}, R_{MN(1)}^{(1)}, R_{MN(2)}^{(2)} \}
\end{equation}
\begin{equation}
\mathcal{R}=R_{MN}G^{MN}=R_{\alpha \beta}g^{\alpha \beta}+R_{(1)(1)}g^{(1)(1)}+R_{(2)(2)}g^{(2)(2)}=R+R^{(1)}+R^{(2)}
\end{equation}
The Euler-Lagrange equations are associated from a Lagrangian of the form
\begin{displaymath}
\mathcal{L}=\sqrt{|G|}\mathcal{R}
\end{displaymath}
where $G=|det(G_{MN})|$.
The scalar curvature $\mathcal{R}$ constructed from $G_{MN}$ differs from the Riemannian one because of the contributions of the internal variables $\phi^{(1)},\phi^{(2)}$. So the scalar curvature $\mathcal{R}$ becomes more general than the one in Brans-Dicke theory.
 In our case the Ricci tensors take the form
\begin{equation}
\widetilde {R}_{\alpha \beta}= R_{\alpha \beta} +\Gamma_{\alpha \beta}^{\mu} \phi _{, \mu} \phi^{-1}+\frac{1}{2}  \phi _{, \alpha} \phi _{, \beta} \phi^{-1}- \phi _{, \alpha \beta} \phi^{-1}
\end{equation}
\begin{equation}
R_{(1)(1)}= R_{(2)(2)}=\frac{1}{2}g^{\alpha \mu} g^{\beta \nu} g_{\alpha \beta ,\mu} \phi_{,\nu}-\frac{1}{2}g^{\mu \nu} (\phi_{,\mu \nu} + \frac{1}{2}g_{,\mu} \phi_{,\nu})
\end{equation}
where $R_{\alpha \beta}$ is the Riemannian Ricci tensor,the indices ($\alpha,\beta,...)$ run over (0,...,3) and (1),(2) represent the fibres.In this model the Langrangian density is given by \cite{stik}
\begin{equation}
\mathcal{L} = \sqrt{|g|} ( \phi R - 2g^{ \alpha \beta} \phi_{; \alpha \beta} + \frac{1}{4} g^{\alpha \beta} \phi_{,\alpha} \phi_{, \beta} / \phi) 
\end{equation}
From (87) we notice that the first and third term of the Lagrangian are similar with the Brans-Dicke theory.In this form of the Lagrangian $\mathcal{L}$ a non-minimal coupling with the scalar field is presented
Varying the action($\mathcal{L}$) with respect to $g_{\mu \nu}$ and $\phi$ we obtain the equations
\begin{equation}
\frac{\delta \mathcal{L}}{\delta g_{\kappa \lambda}} \equiv \left( R^{\kappa \lambda} - \frac{1}{2}g^{\kappa \lambda}R \right) + \frac{1}{4} \left( g^{\alpha \beta} g^{\kappa \lambda} \phi^{-1} - g^{\alpha \kappa} g^{\beta \lambda} \right) \phi_{, \alpha} \phi_{, \beta} + \phi _{| \alpha \beta} (g^{\alpha \kappa} g^{\beta \lambda} - 
g^{\alpha \beta} g^{\kappa \lambda}) = 0
\end{equation}
\begin{equation}
\mathcal{L}_{\phi}=\frac{\delta \mathcal{L}}{ \delta \phi} \equiv  \phi R + \frac{1}{2}g^{\alpha \beta} \phi_{,\alpha} \phi_{,\beta} \phi^{-1} - g^{\alpha \beta} \phi_{; \alpha \beta} = 0
\end{equation}
where the symbol ";" means covariant derivative. We have to note here that despite the fact that we have two scalar fields $\phi^{(1)},\phi{(2)}$ both of them are related with the equation $\phi(x)$ rel.(79). Consequently, the field equation of $\phi$ is taken from (89).
From the Euler-Lagrange equations for the Lagrangian $\mathcal{L}$ of rel.(87) with a potential $V(\phi,\phi^{(1)},\phi^{(2)})$ we get the Klein-Gordon equation which expresses the dynamic of $\phi$ in the following form
\begin{equation}
\left(1+\frac{1}{2} \phi^{-1} \right) \Box \phi -\frac{1}{2} \phi^{-1} g^{\alpha \beta} \phi_{,\alpha}\phi_{,\beta}-\phi R+V'(\phi,\phi^{(1)},\phi^{(2)})=0,
\end{equation}
where $V'=\frac{dV}{d \phi}$.

\subsection{The Raychaudhuri equation of the model}
If we substitute the relations (85)(86) in this model we obtain the Raychaudhuri equation\\

$X^{N}\widetilde{\Theta} |_{N}=\frac{d \tilde{\Theta}}{d\tau}=R_{MN}X^{M}X^{N} - \frac{1}{3} \tilde{\Theta}^{2} - \sigma_{K}^{N} \sigma_{N}^{K}
- \omega_{K}^{N} \omega_{N}^{K} $\\
or
$X^{N}\tilde{\Theta} |_{N}=$\\

$=\{R_{\alpha \beta} +\Gamma_{\alpha \beta}^{\mu} \phi _{, \mu} \phi^{-1}+\frac{1}{2}  \phi _{, \alpha} \phi _{, \beta} \phi^{-1}- \phi _{, \alpha \beta} \phi^{-1}
 + g^{\alpha \mu} g^{\beta \nu} g_{\alpha \beta ,\mu} \phi_{,\nu}-\frac{1}{2}g^{\mu \nu} (\phi_{,\mu \nu} + \frac{1}{2}g_{,\mu} \phi_{,\nu})\} X^{M}X^{N}$\\

\begin{equation}
- \frac{1}{3} \widetilde{\Theta}^{2}- \sigma_{K}^{N} \sigma_{N}^{K} - \omega_{K}^{N} \omega_{N}^{K} \hspace{1cm} 
\end{equation}
where we have considered the conditions
$T_{mk}^{i} = R_{mk}^{b} = 0$ and $R_{MN}$ are given by (82). We can apply the metric of FRW geometry with coordinates $(t,r,\theta, \phi)= (x^{a})$ for the space-time $M \times \phi^{(1)} \times \phi^{(2)}$.We use the Ricci tensors
\begin{displaymath}
R_{00} = 3 \frac{\ddot{a}}{a}=4 \pi G(\mu +3p)
\end{displaymath}
\begin{displaymath}
R_{11} = - \frac{a \ddot{a} + 2\dot{a}^{2} + 2k}{1-kr^{2}}
\end{displaymath}
\begin{displaymath}
R_{22} = -\frac{(a \ddot{a} + 2\dot{a}^{2} + 2k)}{r^{2}}
\end{displaymath}
\begin{displaymath}
R_{33} = -\frac{(a \ddot{a} + 2\dot{a}^{2} + 2k)}{r^{2} \sin^{2} \theta},
\end{displaymath}

\begin{displaymath}
\widetilde{R}_{MN}=R_{00}+R_{11}+R_{22}+R_{33}+R_{(1)(1)}+R_{(2)(2)}=
\end{displaymath}
\begin{displaymath}
=4 \pi G(\mu +3p)+\frac{1}{2}(\phi_{,t})^{2}\phi^{-1}-\phi_{,tt}\phi^{-1}+\frac{a\ddot{a} +2 \dot{a}^{2}+2k}{1-kr^{2}}+\frac{k}{2}(1-kr)^{-1}\phi_{,r}\phi^{-1}+\frac{1}{2}\phi_{,r}^{2}\phi^{-1}-\phi_{,rr}\phi^{-1}-
\end{displaymath}
\begin{displaymath}
-\frac{a\ddot{a} +2 \dot{a}^{2}+2k}{r^{2}} +\frac{1}{2}\phi_{,\theta}^{2}\phi^{-1}-\phi_{,\theta \theta}\phi^{-1} -\frac{a\ddot{a} +2 \dot{a}^{2}+2k}{r^{2}\sin^{2} \theta} +\frac{1}{2} \phi_{,\phi}^{2} \phi^{-1}-\phi_{,\phi \phi} \phi^{-1}+\phi_{,tt}+a^{-2}k \phi_{,r}-
\end{displaymath}
\begin{equation}
-\frac{1-kr^{2}}{a^{4}} \phi_{,rr} +\frac{(4-2kr)r^{3}a^{6}\sin^{2}\theta}{(1-kr)^{2}} \phi_{,r}-(ar)^{-2}\phi_{,\theta \theta}+(\frac{r^{4}a^{6}\sin^{2}\theta}{1-kr^{2}})_{, \theta} \phi_{, \theta}(ar)^{-2} - (ar\sin \theta)^{-2} \phi^{,\phi \phi} 
\end{equation}

From (91) and (92) the tidal force field with $\tilde{R}_{MN}X^{M}X^{N}$ is written
\begin{equation}
\widetilde{R}_{MN}X^{M}X^{N} = \widetilde{\mathcal{A}}_{MN} X^{M}X^{N} 
\end{equation}

where we put
\begin{displaymath}
\tilde{\mathcal{A}}_{MN}=[4 \pi G(\mu +3p)+\frac{1}{2}(\phi_{,t})^{2}\phi^{-1}-\phi_{,tt}\phi^{-1}+\frac{a\ddot{a} +2 \dot{a}^{2}+2k}{1-kr^{2}}+ \frac{k}{2} (1-kr)^{-1} \phi_{,r}\phi^{-1}+\frac{1}{2}\phi_{,r}^{2}\phi^{-1}-\phi_{,rr}\phi^{-1}-
\end{displaymath}
\begin{displaymath}
-\frac{a\ddot{a} +2 \dot{a}^{2}+2k}{r^{2}}+\frac{1}{2}\phi_{,\theta}^{2}\phi^{-1}-\phi_{,\theta \theta}\phi^{-1}-\frac{a\ddot{a} +2 \dot{a}^{2}+2k}{r^{2}\sin^{2} \theta}+\frac{1}{2} \phi_{,\phi}^{2} \phi^{-1}-\phi_{,\phi \phi} \phi^{-1}+\phi_{,tt}+a^{-2}k \phi_{,r}-
\end{displaymath}
\begin{equation}
-\frac{1-kr^{2}}{a^{4}} \phi_{,rr}+\frac{(4-2kr)r^{3}a^{6}\sin^{2}\theta}{(1-kr)^{2}} \phi_{,r}-(ar)^{-2}\phi_{,\theta \theta}+(\frac{r^{4}a^{6}\sin^{2}\theta}{1-kr^{2}})_{, \theta} \phi_{, \theta}(ar)^{-2} -(ar\sin \theta)^{-2} \phi^{,\phi \phi}]
\end{equation}
Therefore the Raychaudhuri equation takes the form
\begin{equation}
X^{N}\widetilde{\Theta} |_{N}=\frac{d \widetilde{\Theta}}{d\tau}=\widetilde{\mathcal{A}}_{MN}X^{M}X^{N} - \frac{1}{3} \widetilde{\Theta}^{2} - \sigma_{K}^{N} \sigma_{N}^{K}
- \omega_{K}^{N} \omega_{N}^{K} 
\end{equation}
If $\widetilde{\mathcal{A}}_{MN}> 0$, it contributes to defocusing or acceleration,whereas if $\widetilde{\mathcal{A}}_{MN}< 0$, we have focusing or deceleration to the evolution of the universe.

\subsection{Contribution of non-linear connection to the equation}
In a more general case, if we consider the torsion (rel(6)) we obtain the Ricci tensors $ K_{\alpha \beta}, K_{(1)(1)}, K_{(2)(2)}$ :

\begin{equation}
K_{\alpha \beta}=\widetilde{R}_{\alpha \beta} +\frac{1}{2}\phi_{,\alpha}\phi^{-1} \left(\frac{\partial{N_{\beta}^{(1)}}}{\partial{\phi^{(1)}}}+\frac{\partial{N_{\beta}^{(2)}}}{\partial{\phi^{(2)}}} \right)
\end{equation}
\begin{equation}
K_{(1)(1)}=R_{(1)(1)} +\frac{1}{2}g^{\mu\nu}\phi_{,\nu}(\frac{\partial{N_{\mu}^{(1)}}}{\partial{\phi^{(1)}}})
\end{equation}
\begin{equation}
K_{(2)(2)}=R_{(2)(2)} +\frac{1}{2}g^{\mu\nu}\phi_{,\nu}(\frac{\partial{N_{\mu}^{(2)}}}{\partial{\phi^{(2)}}})
\end{equation}

where $\widetilde{R}_{\alpha \beta },R_{(1)(1)},R_{(2)(2)}$
are given by (85),(86).\\

The non-linear connections $N_{\beta}^{(1)},N_{\beta}^{(2)}$ can physically represent the interaction field between space-time and inflaton or anisotropy. In addition,the terms $\frac{\partial N_{a}^{(1)}}{\partial \phi^{(1)}},\frac{\partial N_{a}^{(2)}}{\partial \phi^{(2)}}$ can be interpreted as the variation of the interactive fields with respect to local anisotropy of space-time or the inflaton.In this case the field equations produced by the variational principle of the generalized Lagrangian \cite{stik}
\begin{displaymath}
\widetilde{\mathcal{L}}= \sqrt{|G|} G^{AB} K_{AB}
\end{displaymath}
given by 
\begin{equation}
\delta \widetilde{\mathcal{L}}_{\phi}=\frac{\delta \widetilde{\mathcal{L}}}{\delta \phi}= \sqrt{|g|} g^{\alpha \beta} \left( R_{\alpha \beta} - \frac{\phi_{; \alpha \beta}}{\phi} + \frac{\phi_{; \alpha} \phi_{; \beta}}{2 \phi^{2}} \right)  - \frac{\delta}{\delta x^{a}} \left[ \sqrt{|g|} g^{\alpha \beta} \left( \frac{\partial N_{\beta}^{(1)}}{\partial \phi^{(1)}} + \frac{\partial N_{\beta}^{(2)}}{\partial \phi^{(2)}}\right) \right]=0
\end{equation}
where $\mathcal{K}_{AB} = \{ K_{\alpha \beta}, K_{(1)(1)}, K_{(2)(2)} \}.$

In this structure of spacetime the Raychaudhuri equation is written as
\begin{equation}
X^{N}\widetilde{\Theta} |_{N}=\frac{d \widetilde{\Theta}}{d\tau}=\mathcal{K}_{LN} X^{L}X^{N} - \frac{1}{3} \widetilde{\Theta}^{2} - \sigma_{K}^{N} \sigma_{N}^{K}
- \omega_{K}^{N} \omega_{N}^{K}
\end{equation}
Following through the same procedure as above we can obtain the Raychaudhuri equations with the contribution of non-linear connection.

\begin{equation}
X^{N}\widetilde{\Theta} |_{N}=\frac{d \widetilde{\Theta}}{d\tau}=\widetilde{\mathcal{B}}_{MN} X^{M}X^{N} - \frac{1}{3} \widetilde{\Theta}^{2} - \sigma_{K}^{N} \sigma_{N}^{K}-\omega_{K}^{N} \omega_{N}^{K}
\end{equation}

where 
\begin{equation}
\tilde{\mathcal{B}}_{MN}=\tilde{\mathcal{A}}_{MN}+\frac{1}{2}\phi_{,\alpha}\phi^{-1}\left(\frac{\partial{N_{\beta}^{(1)}}}{\partial{\phi^{(1)}}}+\frac{\partial{N_{\beta}^{(2)}}}{\partial{\phi^{(2)}}}\right)+\frac{1}{2}g^{\mu\nu}\phi_{,\nu}\left(\frac{\partial{N_{\mu}^{(1)}}}{\partial{\phi^{(1)}}}\right)+\frac{1}{2}g^{\mu\nu}\phi_{,\nu}\left(\frac{\partial{N_{\mu}^{(2)}}}{\partial{\phi^{(2)}}}\right) 
\end{equation}
and $\tilde{\mathcal{A}}_{MN}$ is given by (94). From (101),(102) it seems that the interaction terms can contribute to decelerate form of space-time.

\section{Conclusions}

In the present work, the fundamental role of Raychaudhuri equation for the general relativity is extended in the framework of generalized geometrical structures of a Finsler-Randers spacetime and of generalized scalar-tensor theories. Additional terms in the equation are introduced because of the anisotropic curvatures and of the form of geodesics (rel.(59)). In this approach we used the Cartan connection, since it preserves the norm of a vector (time-like,null) under a parallel propagation and it is convenient for studying of modified gravitational and cosmological theories. The Raychaudhuri equation was also derived in connection with the FRW model for the FR space-time. In par.3 we studied the energy conditions for the FR-cosmology and their relations with the FRW-cosmology. In addition we provide the bounce conditions for a FR cosmology. It was proved that in a bounce both the energy conditions are identified. 
Our work is based on the geometry of the bundle consisting of a background manifold and two scalars as fibres.In this framework, we gave the form of the Klein-Gordon equation and we studied the Raychaudhuri equation in a generalized form of scalar-tensor theory of a model $M \times \{ \phi^{(1)} \} \times \{ \phi^{(2)} \}$ with the presence of non-linear connections.These connections as extra terms can play a significant role to the gravitational influence and its interaction with other fields.This means that these additional terms/fields will differentiate the evolution of accelerated expansion of the universe(focusing/defocusing) as it is obvious from the rel.(93),(94),(101),(102).

\section{Appendix A}

The Lagrangian metric function in a Finsler-Randers(FR) space is given by
\begin{displaymath}
L(x,y)=(g_{ij}(x)y^{i}y^{j})^{1/2} + a_{i}(x)y^{i}
\end{displaymath}
where $a_{i}$ represents a covector and $g_{ij}$ the metric tensor of the pseudo-Riemannian spacetime. By using Cartan's connection the deformation tensor$\Delta^{i}_{jk}$ in a FR space is defined as 
\begin{displaymath}
\Delta^{i}_{jk}=\Gamma_{jk}^{i}-L_{jk}^{i}, \hspace{9,5cm} (A.1)
\end{displaymath}
 \cite{matsu},\cite{yasuda},\cite{st3} where $\Gamma_{jk}^{}$ are the Christoffel symbols of Riemannian space and $L_{jk}^{i}$ the Cartan connections coefficients of FR space.
The contraction terms $\Delta^{i}_{j0}, \Delta_{00}^{i}$ are given by
\begin{displaymath}
\Delta_{j0}^{i}=p^{i}a_{(j0)}+ \frac{1}{2} h_{j}^{i} a_{00} + G^{is}(a_{[sj]}+a_{[s0]}p_{j})-2g^{ms}A_{jm}^{i}a_{[s0]}/\tau,   \hspace{2cm}  (A.2)
\end{displaymath}
\begin{displaymath}
\Delta_{00}^{i}=2( \lambda p^{i}+g^{is}a_{[s0]}/ \tau),  \hspace{8,2cm} (A.3)
\end{displaymath}
where the torsion tensor $A_{jk}^{i}$ is defined by
\begin{displaymath}
A_{jk}^{i}=(h_{j}^{i}L_{k}+h_{k}^{i}L_{j}+\frac{2}{5} h_{jk} A^{i})/2,
\end{displaymath}
and 
\begin{displaymath}
a_{ij}= \triangledown_{j}a_{i},
\end{displaymath}
with $\triangledown_{j}$ we denote  the covariant derivative with respect to the Riemannian space.Also the symbols ( ),[ ], mean symmetrization, antisymmetrization and "o" represents contraction with $p^{i}$
\begin{displaymath}
h_{ij}= \tau(g_{ij}-a_{i}a_{j})
\end{displaymath}
\begin{displaymath}
h_{j}^{i}=G^{ik} h_{kj}=\delta_{j}^{i}-p^{i}p_{j},
\end{displaymath}
$G^{ij}$ denote  the components of the inverse metric of FR space and
\begin{displaymath}
\tau= L/g^{1/2}
\end{displaymath}
\begin{displaymath}
p^{i}=y^{i}/L
\end{displaymath}
\begin{displaymath}
p_{i}=\partial L/ \partial Y^{i}
\end{displaymath}
\begin{displaymath}
L_{i}=a_{i}-\mu l_{i}
\end{displaymath}
\begin{displaymath}
\mu=a_{i}l^{i}
\end{displaymath}
\begin{displaymath}
\ell^{i}=y^{i}/g^{1/2}
\end{displaymath}

If $\Delta_{jk}^{i}=0$ from rel.(A.1), $L_{jk}^{i}$ are the Christoffel symbols $\Gamma_{jk}^{i}$ of the Riemannian space. In this case the Randers space is called a Landsberg-Randers space. The h-curvature $\widetilde{R}_{hjk}^{i}$ of FR space is connected with the Riemannian one $R_{hjk}^{i}$ with the relation
\begin{displaymath}
\widetilde{R}_{hjk}^{i}=R_{hjk}^{i}+A_{hr}^{i}R_{0jk}^{r},  \hspace{2cm} (A.4)
\end{displaymath}
In addition if the torsion tensor $A_{hr}^{i}=0$ the Finsler Randers spaces reduces in a Riemannian one.

\end{document}